\documentclass[journal]{IEEEtran}
\usepackage{ifpdf}
\ifpdf
\else
\fi

\usepackage{cite}

\ifCLASSINFOpdf
  \usepackage[pdftex]{graphicx}
  \graphicspath{{../pdf/}{../jpeg/}}
  \DeclareGraphicsExtensions{.pdf,.jpeg,.png}
\else
  \usepackage[dvips]{graphicx}
  \graphicspath{{../eps/}}
  \DeclareGraphicsExtensions{.eps}
\fi
\usepackage{amsmath} 
\usepackage{amssymb}  
\usepackage{mathptmx} 

\usepackage{algorithmic}

\usepackage{array}

\usepackage{mdwmath}
\usepackage{mdwtab}

\usepackage{eqparbox}

\ifCLASSOPTIONcompsoc
\usepackage[tight,normalsize,sf,SF]{subfigure}
\else
\usepackage[tight,footnotesize]{subfigure}
\fi

\usepackage[caption=false,font=footnotesize,tight]{subfig}
\usepackage{fixltx2e}

\usepackage{stfloats}

\usepackage{float}

\usepackage{url}

\begin{document}
%
\title{Computationally efficient formulation of relay
operator for Preisach hysteresis modeling}
%
%
%

\author{\IEEEauthorblockN{
        Michael~Ruderman,~\IEEEmembership{Member,~IEEE}}\\\vspace{0.5cm}
\thanks{ Correspondence: M. Ruderman (e-mail: ruderman@vos.nagaokaut.ac.jp)}
\authorblockA{Nagaoka University of Technology, Department of Electrical Engineering, Nagaoka, 940-2188, Japan} }

%
%

\markboth{IEEE TRANSACTIONS ON MAGNETICS}%
{Shell \MakeLowercase{\textit{et al.}}: Bare Demo of IEEEtran.cls
for Journals}
%



\maketitle

\begin{abstract}
\boldmath An algebraic expression for the Preisach hysteron, which
is a non-ideal (delayed) relay operator, is formulated for a
computationally efficient real-time implementation. This allows
representing the classical scalar Preisach hysteresis model as a
summation of a large number of weighted hysterons which
computation can be accomplished in parallel. The latter makes
possible an efficient FPGA or ASIC realization of the scalar
Preisach hysteresis model that can be useful for multiple
applications. The signal flow which specifies the model
implementation is provided in form of the block diagram. The
proposed computation of Preisach hysterons, aggregated to the
entire Preisach hysteresis model, is evaluated numerically and on
a real-time hardware platform.
\end{abstract}

\begin{IEEEkeywords}
Hysteresis, Preisach model, non-ideal relay, real-time
computation, hysteron, nonlinear operator
\end{IEEEkeywords}

%

\IEEEpeerreviewmaketitle

\bstctlcite{references:BSTcontrol}

\section{INTRODUCTION}
\label{sec:1}

The scalar Preisach hysteresis (SPH) model is a universal mean for
describing the rate-independent hysteresis phenomena with
congruent loops and erasable, i.e. with wiping out property,
memory, see e.g. \cite{Maye03} for details. The operator-based
mathematical formalism of Preisach hysteresis model has been
thoroughly studied and described in seminal works
\cite{Maye03,Krejci96,Visit94,BrokSprek96}. The definition of a
Preisach hysteresis operator is general enough and a variety of
formulations and implementations have been proposed in the
literature, often depending on the application at hand and
particular computation method yielding the hysteresis state and
correspondingly output. A very common form of the SPH model,
suitable for the fast real-time inversion and control, are the
so-called Everett integrals, as used e.g. in works
\cite{davino2005,davino2008,davino2014}. This form relies on the
stored Everett integrals matrix, which is the measured realization
of first-order reversal curves, and computes the hysteresis output
by an explicit formula provided in \cite{doong1985,mayergoyz1988}.
Using the space representation of Preisach plane, which is a
well-known geometric interpretation of the Preisach model, and
corresponding memory-line interface several schemes evaluate the
output increment of the Preisach operator, see e.g.
\cite{mccarthy2011}. A related attempt of a continuous-time
state-space formulation of the SPH model has been made in
\cite{RudBert11} and applied, with discrete-time, in the control
in \cite{ruderman2014}.

At the same time, the classical SPH model formulation, as in
\cite{Maye86}, is given by
\begin{equation}\label{1}
    f(t) = \iint \limits_{\alpha \geq \beta} \mu (\alpha, \beta)
    \hat{\gamma} _{\alpha \beta} x (t) d\alpha d\beta ,
\end{equation}
where $x(t)$ is a piecewise-monotone input function and $\mu
(\alpha, \beta)$ is an integrable nonnegative function defined
over the $(\alpha, \beta)$ Preisach (half) plane with $\alpha \geq
\beta$. The Preisach weighting function $\mu(\cdot)$, also denoted
as Preisach density function, has a finite support (on the
Preisach plane) within some triangle $T$ which is given by the
hypotenuse $\alpha = \beta$ and the right angle at $(x_{\max},
x_{\min})$. Therefore, for a given input domain $[x_{\min}, ...,
x_{\max}]$ the Preisach density function $\mu (\alpha, \beta)$
with $(\alpha, \beta) \in T$ entirely parameterizes the Preisach
hysteresis operator. The overall state of the Preisach hysteresis
is determined by an infinite set of Preisach hysterons
$\hat{\gamma}_{\alpha\beta}$ which are the non-ideal relays with
two discrete states $\hat{\gamma}_{\alpha\beta} \in \{-1, +1\}$.
Further mathematical properties of the Preisach hysteresis
operator can be found e.g. in the works \cite{Maye86,brokate1989}.

\begin{figure}[!h]
\centering
\includegraphics[width=0.45\columnwidth]{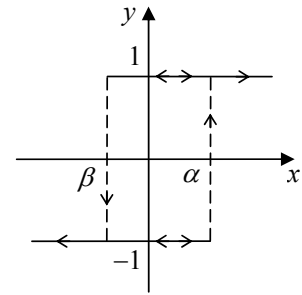}
\caption{State transitions of Preisach hysteron (non-ideal relay)}
\label{fig:1}
\end{figure}

While the discontinuous state transitions of a Preisach hysteron
can be represented as in Fig. \ref{fig:1},
the often used notation of the hysteron's states is given by
\begin{equation}\label{2}
y(t) = \hat{\gamma} _{\alpha\beta} x(t) =
\left\{%
\begin{array}{ll}
    -1, & \hbox{if } x(t) \leq \beta, \\
    +1, & \hbox{if } x(t) \geq \alpha, \\
    y(t_{0}), & \hbox{if } \beta < x(\tau) < \alpha \quad \forall \quad \tau \in[t_{0},
    t].\\
\end{array}%
\right.
\end{equation}
Therefore, the state of each hysteron is uniquely determined
provided the initial sate at the time instant $t_{0}$ is given.
From the practical, and thus application-related, point of view a
computationally efficient and possibly real-time compatible
realization of the total Preisach operator, and correspondingly
single Preisach hysterons if required, can be a crucial factor
when using the SPH model. Also in view of a finite memory
implementation, when a non-parametric measure on the Preisach
plane is assumed, an efficient storage of the hysteresis
parameters is desirable along with the computation of hysteresis
states. Recall that a non-parametric Preisach measure does not
assume any analytic form of the Preisach density function and
implements a discrete mesh (also denoted as grid) on the Preisach
plane. A comparison of parametric and non-parametric
identification of the Preisach hysteresis can be found in
\cite{HenzRuck02}.

Apart from a fast real-time formulation of the SPH model and its
inverse, reported e.g. in works
\cite{davino2005,davino2008,davino2014}, a parallelizm in
processing (computing) a large number of elementar hysteresis
operators, e.g. hysterons, can be highly efficient for multiple
applications. Here the FPGA (Field Programmable Gate Array) and
ASIC (Application-Specific Integrated Circuit) hardware solutions
appear as well-promising when implementing a hysteresis model.
This is quite naturale when neither sequential processor-based
computations are required. The FPGA-based solutions of hysteresis
computation and correspondingly control have been shown in
\cite{tan2008,iyer2009} for Krasnosel'skii-Pokrovskii and in
\cite{Janocha2008} for Prandtl-Ishlinskii hysteresis operators.
Among the real-time applications, the work \cite{Liu2014} can be
mentioned here, while the developed and VHDL (Very High Speed
Integrated Circuit Hardware Description Language) implemented
nonlinear hardware model of the power transformer included a
real-time hysteresis computation.

The aim of this paper is to propose a computationally efficient
formulation of a relay operator, i.e. Preisach hysteron, and
thereupon based implementation of the SPH model. The main
contribution is in deriving a rigorous algebraic form for (2)
which allows a real-time execution of arbitrary input series and,
above all, parallel processing of multiple hysterons. A low set of
primitive operations required per hysteron -- two summations, two
sign operators, and two comparisons, plus one memory storage of
the previous state -- allow a FPGA-compatible implementation
without any complex programm routines. In the following, we
describe in details the proposed formulation of the relay operator
(2) and comment on the related entire SPH model in Section II. In
Section III, a numerical evaluation is shown along with a hardware
validation of the real-time computation. The latter is
accomplished, in the first stage insofar, on the dSpace DSP
(digital signal processor) platform at 2 kHz sampling rate. A
relatively high but also limited, through the DSP operation memory
capacity, number of hysterons $N=210$ is assumed. The paper is
concluded by a brief summery and discussion both provided in
Section IV.

\section{COMPUTATIONALLY EFFICIENT FORMULATION OF RELAY OPERATOR}
\label{sec:2}

In order to derive a computationally efficient algebraic
expression for the non-ideal relay (further denoted as hysteron),
as in Fig. \ref{fig:1}, consider first the memoryless sign
nonlinearities $\mathrm{sign}(x-\beta)$ and
$\mathrm{sign}(x-\alpha)$ with $\alpha
> \beta$. Both represent the one-way switching behavior of a
hysteron at thresholds $\beta$ and $\alpha$. The first one is for
a monotonically decreasing input, i.e. $dx/dt < 0$, and the second
one is for a monotonically increasing input, i.e. $dx/dt >0$.
\begin{figure}[!h]
\centering
\includegraphics[width=0.45\columnwidth]{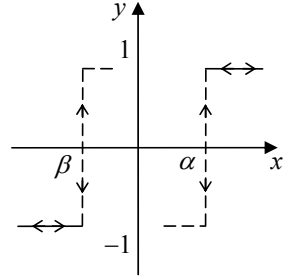}
\caption{State transitions of $\mathrm{sign}$ operators at
hysteron's thresholds} \label{fig:2}
\end{figure}
When mapping both sign operators into the $(x, y)$ state diagram,
see Fig. \ref{fig:2}, one can recognize that the minimal state of
both, i.e.
$$
\min\left[\mathrm{sign}(x-\beta), \mathrm{sign}(x-\alpha)\right],
$$
describes the hysteron's behavior on the interval $x \in (-\infty,
\beta) \vee (\alpha, \infty)$. For the interval between the
threshold values, i.e. $ x \in [\beta, \alpha]$, the operator
state remains first undefined.

\begin{figure}[!h]
\centering
\includegraphics[width=0.45\columnwidth]{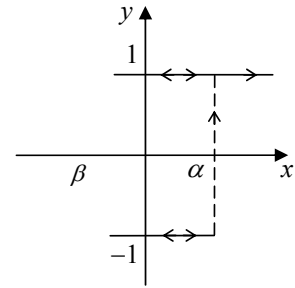}
\caption{State transitions of $\max \left[y(t_{-}),
\mathrm{sign}(x-\alpha) \right]$ for $x \in (\beta,\infty)$}
\label{fig:3}
\end{figure}

Further consider the interval $ x\in (\beta, \infty)$ for which
the possible hysteron transitions can be represented as in Fig.
\ref{fig:3}.
Here one can see that the recent state $y$ depends not only on the
recent input but equally on the previous state denoted by
$y(t_{-})$. Since the switching condition is captured by
$\mathrm{sign}(x-\alpha)$, the latter is to combine with the
previous state, and that by the maximal value selection
\begin{equation}\label{3}
\max\left[y(t_{-}), \mathrm{sign} (x-\alpha)\right].
\end{equation}

\begin{figure}[!h]
\centering
\includegraphics[width=0.85\columnwidth]{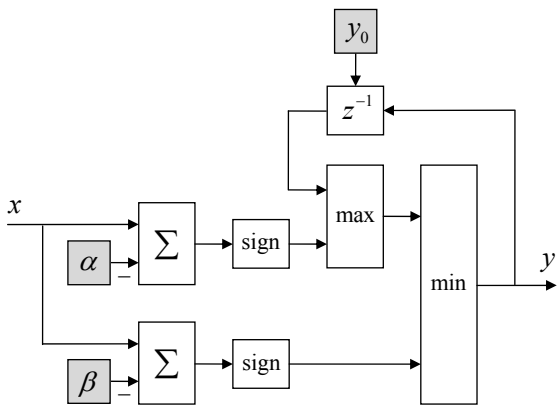}
\caption{Signal flow diagram of Preisach hysteron (non-ideal
relay)} \label{fig:4}
\end{figure}

Now one can recognize that in order to complete the overall set of
hysteron transitions, and correspondingly states,  the subset (3)
has to be combined with the $\mathrm{sign}(x-\beta)$ switch, here
again by the minimal value selection, so that
\begin{equation}\label{4}
y(t)=\min\left[\mathrm{sign}(x-\beta), \max \left[ y(t_{-}),
\mathrm{sign} (x-\alpha) \right] \right].
\end{equation}
Introducing the initial state
\begin{equation}\label{5}
y(t_{0})= \left\{%
\begin{array}{ll}
    \mathrm{sign}(x(t_{0})), & \hbox{ if } x(t_{0}) \in (-\infty, \beta) \vee (\alpha, \infty), \\
    \left[-1,+1\right], & \hbox{ otherwise.}
\end{array}%
\right.
\end{equation}
we obtain the overall input-output behavior of a hysteron, and
that in the closed analytic form (4), (5). In view of a possible
real-time implementation, the related signal flow can be
represented by the block diagram as in Fig. \ref{fig:4}. One can
see that the proposed implementation requires solely 2 summation
operators, 2 sign operators, and two comparators (max, min), plus
an additional memory block for storage of the previous output
state. For the parameterization of hysteron, the $\alpha$,
$\beta$, and $y_{0}$ values have to be stored. Further we note
that in terms of incorporating the single hysterons into the
Preisach hysteresis model each output $y_{i}$, of the overall $i
\in N$ hysterons, is subject to an additional gain $W_{i}$. The
latter is directly applicable at the output signal flow of Fig.
\ref{fig:4}.

Most important is the fact that the real-time computation of
multiple hysterons does not require any sequentially executed
programm codes. Therefore a fully parallel (one-step) computation
of a large number of single hysterons can be realized by means of
e.g. FPGAs or ASICs, developed for a particular application. One
can see that the summation
\begin{equation}\label{6}
f(t)=\sum \limits_{i=1}^{N} W_{i} y_{i}\left(x(t), \alpha_{i},
\beta_{i}, y_{i}(t_{0})\right)
\end{equation}
of overall $N$ weighted hysterons, all connected to the same input
channel and executed in parallel, provides the entire Preisach
hysteresis model. The latter is parameterized by the vectors of
threshold values $\alpha, \beta \in \mathbb{R}^{N\times1}$ with
$\alpha_{i} \geq \beta _{i}$, initial states $y(t_{0})\in
\mathbb{R}^{N\times1}$ with $y_{i}(t_{0}) \in [-1, +1]$, and
hysteron weights $W \in \mathbb{R}^{N\times1}$ with $W_{i}
> 0$. Furthermore, it is worth noting that the parallel real-time computation
according to Fig. 4 and eq. (4) allows equally for further
advanced features of the Preisach hysteresis model, like e.g. the
direct recursive identification \cite{ruderman2013,ruder2015}.
These are, however, out of scope of the recent communication and
will be addressed in the future works.

\section{NUMERICAL AND HARDWARE EVALUATION}
\label{sec:3}

The computation of SPH model, and that according to (4)--(6), is
first evaluated within a numerical simulation (MATLAB/Simulink
R2010b). A $80 \times 80$ mesh on the discretized Preisach plane
is assumed which results in a total of 3240 hysterons. An uniform
Preisach density function is assumed, for the sake of simplicity,
that means the single weights of all hysterons have the same
value. The latter is set so that the total co-domain is $f(x) \in
[-1, ...,1]$. A monotonically decreasing sinusoidal input has been
applied which results in a set of the nested minor hysteresis
loops running towards the $(x,f)$ origin. The hysteresis loops
recorded from the numerical simulation are shown in Fig. 5.

As next, we evaluate the real-time SPH model computation,
according to (4)--(6), on the available DSP hardware platform
dSpace DS1104CLP. Note that the latter does not provide a parallel
processing of single hysterons but allows to prove the correctness
of a real-time execution of hysteron's implementation as in Fig.
4. The sampling rate is set to 2 kHz and the assumed number of
hysterons is 210. The latter is limited by the available DSP
operation memory and embedded compiler. First, a 1Hz sinusoidal
input is proceeded during the 120 sec runtime. The recorded 120
major hysteresis loops are shown in Fig. 6 (a) over each other.
Second, the white noise input which is low-pass (10 Hz cut-off
frequency) filtered is proceeded during the 120 sec runtime. The
recorded hysteresis trajectories are shown in Fig. 6 (b). Note
that due to a large sporadic amplitude variations of the noise
input series, the depicted (overlapping) trajectories are
smoothing the hysteresis output steps, which are naturally related
to the switching of single hysterons and clearly visible in Fig. 6
(a). Apart from that, the grey-shaded hysteresis area, inside of
the major loop in Fig. 6 (b), discloses a correct response of
implemented SPH model to a long-range random input sequence.
\begin{figure}[!h]
\centering
\includegraphics[width=0.95\columnwidth]{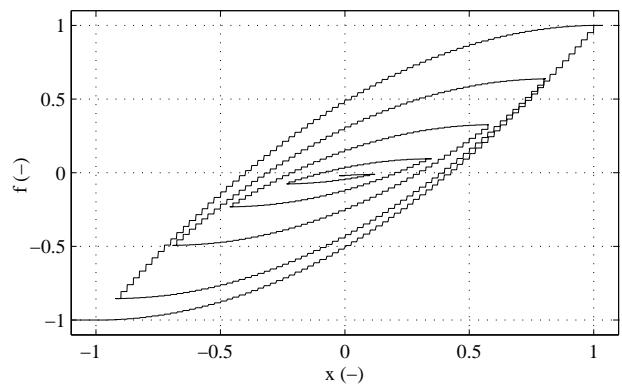}
\caption{Hysteresis loops of scalar Preisach hysteresis model (6)
with 3240 hysterons computed for a monotonically decreasing
sinusoidal input} \label{fig:5}
\end{figure}
\begin{figure}[!h]
\centering
\includegraphics[width=0.49\columnwidth]{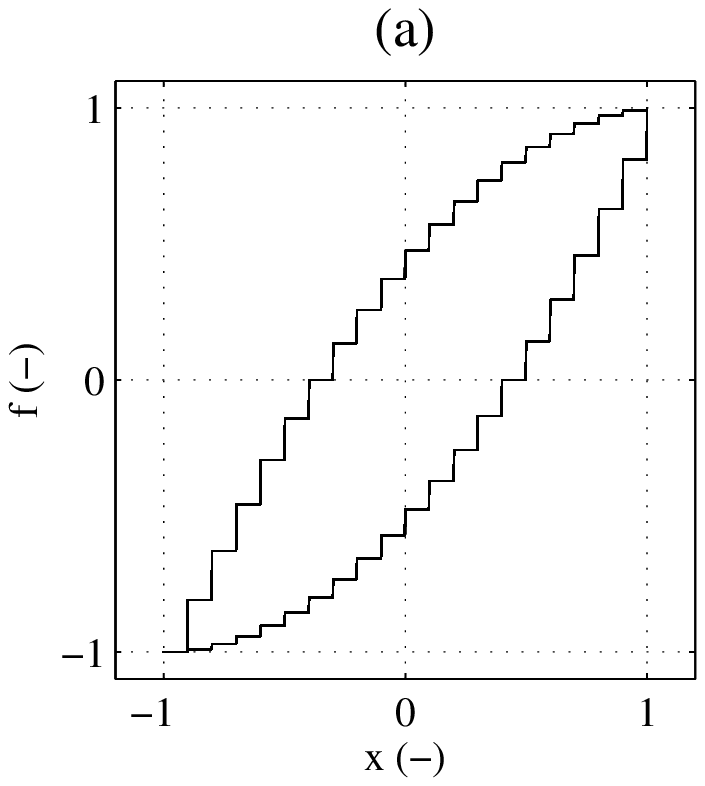}
\includegraphics[width=0.49\columnwidth]{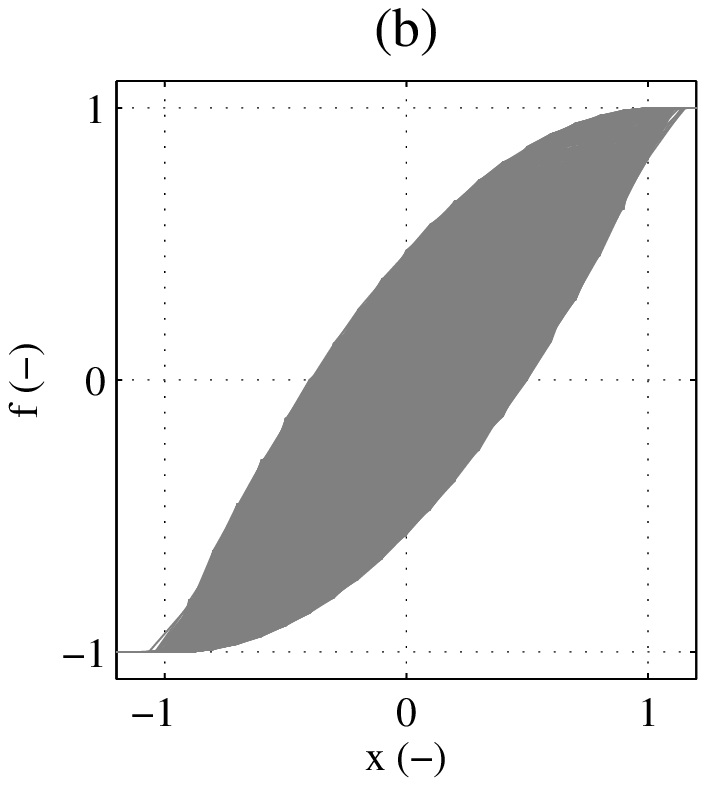}
\caption{Hysteresis loops of scalar Preisach hysteresis model (6)
with 210 hysterons computed on the real-time DSP platform at 2 kHz
sampling rate; (a) 120 major loops at 1 Hz sinusoidal input, (b)
hysteresis trajectories at 10 Hz cut white noise input during 120
sec runtime.} \label{fig:6}
\end{figure}

\section{SUMMARY and DISCUSSION}
\label{sec:4}

\subsection{Summary}

A computationally efficient algebraic form of the Preisach
hysteron, which is a non-ideal relay, has been proposed. This
allows applying a large number of hysterons, all executed
(computed) in parallel, for a real-time implementation of the
scalar Preisach hysteresis model. The proposed formulation
utilizes a low number of primitive operations connected into a
signal flow chart without any complex computational routines. This
allows a direct FPGA or ASIC based hardware implementation which
can be efficient for multiple applications. The proposed
computation of hysterons, and thereupon assembled SPH model, has
been evaluated in the numerical simulation and, additionally, on
the DSP based real-time hardware. In perspective, the proposed
solution is to be implemented and evaluated on a FPGA board using
the intrinsic parallelizm of computing a large number of
hysterons.

\subsection{Discussion}

Theoretically, the single limitation posed on the usage of
proposed method arises from the memory capacities of the low-end
hardware platform at hand, i.e. FPGA or ASIC. The utilized
$\min$-, $\max$-, and $\mathrm{sign}$-operators constitute the
simple comparators at the low-level. Additionally, two summation
and one time-delay functional blocks are required per each
hysteron unit. The time-delay block is a standard sample-and-hold
element switched in a feedback manner. Therefore, the sampling
time of computing a hysteron, and consequently all hysterons in
parallel, is only double of the base sampling time of the hardware
target. Apart from the summation block, with overall $N$ input
channels as in eq. (6), one common zero constant for the
$\mathrm{sign}$-related comparators and two common initial state
constants $[-1,1]$ have to be recorded in a ROM memory.
Furthermore, three additional constants have to be statically
memorized (stored) per each hysteron block. These are the
threshold and weight parameters.

The proposed method is suitable for a relatively large class of
different applications, whenever a rate-independent Preisach
hysteresis computation in real-time is required. One of the
possible application fields is a hardware in the loop during the
design, optimization, and fault detection or monitoring of the
magnetic devices, see e.g. \cite{RosenEtAl10,ruderman2013b} as
potential examples. Another thinkable application is an online
(recursive) identification of hysteresis behavior in a process,
see e.g. \cite{ruder2015}. Here the proposed method is easily
extendable for a recursive scheme of parameter adaption. Finally
the model-based control applications with hysteretic actuators,
see e.g.
\cite{davino2005,Janocha2008,iyer2009,esbrook2013,ruderman2014},
can generally benefit from the proposed fast computation, while
various strategies of using the hysteresis model in control are
thinkable.

\ifCLASSOPTIONcaptionsoff
  \newpage
\fi



%




\bibliographystyle{IEEEtran}
\bibliography{references}

\end{document}